\documentclass[12pt]{aastex62}

\usepackage{natbib}
\usepackage{enumitem}
\usepackage{amsmath,graphicx}
\usepackage{longtable}
\usepackage{threeparttable}
\usepackage{listings}
\usepackage{hyperref}
\hypersetup{
  linkcolor=red,
  citecolor=green,
  filecolor=cyan,
  urlcolor=magenta,
}

\shorttitle{SDSS-RM photometry}
\shortauthors{Kinemuchi et al.}

\begin{document}

\title{The Sloan Digital Sky Survey Reverberation Mapping Project:
  Photometric $g$ and $i$ Light Curves}

\author{K. Kinemuchi}
\affil{Apache Point Observatory/New Mexico State University, 2001 Apache Point Road, Sunspot, NM 88349}

\author{Patrick B.\ Hall}
\affil{Department of Physics and Astronomy, York University, Toronto,
  Ontario M3J 1P3,  Canada}

\author{Ian McGreer}
\affil{Steward Observatory, University of Arizona, 933 North Cherry
  Avenue, Tucson, AZ 85721-0065, USA}

\author{C.~S.~Kochanek}
\affil{Department of Astronomy, The Ohio State University, 140 West 18th Avenue, Columbus, OH 43210, USA}

\author{Catherine~J.~Grier}
\affiliation{Steward Observatory, The University of Arizona, 933 North Cherry Avenue, Tucson, AZ 85721, USA} 
\affiliation{Department of Astronomy and Astrophysics, Eberly College of Science, The Pennsylvania State University, 525 Davey Laboratory, University Park, PA 16802}
\affiliation{Institute for Gravitation \& the Cosmos, The Pennsylvania State University, University Park, PA 16802}

\author{Jonathan Trump}
\affil{Department of Physics, University of Connecticut, 2152 Hillside
Road, Unit 3046, Storrs, CT 06269, USA}

\author{Yue Shen}
\altaffiliation{Alfred P. Sloan Research Fellow}
\affil{Department of Astronomy, University of Illinois at Urbana-Champaign, Urbana, IL 61801, USA } 
\affil{National Center for Supercomputing Applications, University of Illinois at Urbana-Champaign, Urbana, IL 61801, USA}

\author{W.\ N.\ Brandt}
\affiliation{Department of Astronomy and Astrophysics, Eberly College of Science, The Pennsylvania State University, 525 Davey Laboratory, University Park, PA 16802}
\affiliation{Institute for Gravitation \& the Cosmos, The Pennsylvania State University, University Park, PA 16802}
\affiliation{Department of Physics, The Pennsylvania State University,
  University Park, PA 16802, USA}

\author{W.\ M.\ Wood-Vasey}
\affiliation{Pittsburgh Particle Physics, Astrophysics, and Cosmology Center (PITT PACC).
Physics and Astronomy Department, University of Pittsburgh,
Pittsburgh, PA 15260, USA}

\author{Xiaohui Fan}
\affiliation{Steward Observatory, The University of Arizona, 933 North Cherry Avenue, Tucson, AZ 85721, USA} 

\author{Bradley~M.~Peterson}
\affiliation{Department of Astronomy, The Ohio State University, 140 W 18th Avenue, Columbus, OH 43210, USA}
\affiliation{Center for Cosmology and AstroParticle Physics, The Ohio State University, 191 West Woodruff Avenue, Columbus, OH 43210, USA}
\affiliation{Space Telescope Science Institute, 3700 San Martin Drive, Baltimore, MD 21218, USA }

\author{Donald P. Schneider}
\affiliation{Department of Astronomy and Astrophysics, Eberly College of Science, The Pennsylvania State University, 525 Davey Laboratory, University Park, PA 16802}
\affiliation{Institute for Gravitation \& the Cosmos, The Pennsylvania State University, University Park, PA 16802}

\author{Juan V. Hern{\'a}ndez Santisteban}
\affiliation{SUPA Physics \& Astronomy, University of St. Andrews, Fife, KY16 9SS, Scotland, UK}

\author{Keith Horne}
\affiliation{SUPA Physics \& Astronomy, University of St. Andrews, Fife, KY16 9SS, Scotland, UK}

\author{Yuguang Chen}
\affiliation{California Institute of Technology, 1200 E California Blvd., MC 249-17, Pasadena, CA 91125, USA}  

\author{Sarah~Eftekharzadeh}
\affiliation{Department of Physics and Astronomy, University of Utah, 115 S. 1400 E., Salt Lake City, UT 84112, USA}

\author{Yucheng Guo}
\affiliation{Department of Astronomy, School of Physics, Peking University, Beijing 100871, China} 

\author{Siyao~Jia}
\affiliation{Department of Astronomy, University of California, Berkeley, CA 94720, USA}

\author{Feng Li} 
\affiliation{School of Mathematics and Physics, Changzhou University, Changzhou 213164, China} 

\author{Zefeng Li} 
\affiliation{Department of Astronomy, School of Physics, Peking University, Beijing 100871, China} 

\author{Jundan Nie}
\affiliation{Key Laboratory of Optical Astronomy, National Astronomical Observatories, Chinese Academy of Sciences, Beijing 100012, China}

\author{Kara~A.~Ponder}
\affiliation{Berkeley Center for Cosmological Physics,  University of
  California Berkeley, 341 Campbell Hall, Berkeley, CA 94720, USA  }

\author{Jesse~Rogerson}
\affiliation{Canada Aviation and Space Museum, 11 Aviation Parkway, Ottawa, ON, K1K 4Y5, Canada}
\affiliation{Department of Physics and Astronomy, York University, Toronto, ON M3J 1P3, Canada}

\author{Tianmen Zhang} 
\affiliation{Key Laboratory of Optical Astronomy, National Astronomical Observatories, Chinese Academy of Sciences, Beijing 100012, China}
\affiliation{School of Astronomy and Space Science, University of Chinese Academy of Sciences}

\author{Hu Zou} 
\affiliation{Key Laboratory of Optical Astronomy, National Astronomical Observatories, Chinese Academy of Sciences, Beijing 100012, China}

\author{Linhua Jiang}
\affiliation{Kavli Institute for Astronomy and Astrophysics, Peking University, Beijing 100871, China}

\author{Luis C. Ho}
\affiliation{Kavli Institute for Astronomy and Astrophysics, Peking University, Beijing 100871, China} 
\affiliation{Department of Astronomy, School of Physics, Peking
  University, Beijing 100871, China} 

\author{Jean-Paul Kneib}
\affiliation{Institute of Physics, Laboratory of Astrophysics, Ecole Polytechnique F\'ed\'erale de Lausanne (EPFL), Observatoire de Sauverny, 1290 Versoix, Switzerland}
\affiliation{Aix Marseille Universit\'e, CNRS, LAM (Laboratoire d'Astrophysique de Marseille) UMR 7326, 13388, Marseille, France}

\author{Patrick Petitjean}
\affiliation{Institut d'Astrophysique de Paris, Sorbonne Universit\'e and CNRS,
   98bis Boulevard Arago, 75014, Paris, France}

\author{Nathalie Palanque-Delabrouille}
\affiliation{IRFU, CEA, Universit\'e Paris-Saclay, F-91191 Gif-sur-Yvette, France}

\author{Christophe Yeche}
\affiliation{IRFU, CEA, Universit\'e Paris-Saclay, F-91191 Gif-sur-Yvette, France}

\begin{abstract}
The Sloan Digital Sky Survey Reverberation Mapping (SDSS-RM) program
monitors 849 active galactic nuclei (AGN) both spectroscopically and photometrically.
The photometric observations used in this work span over four years and provide an
excellent baseline for variability studies of these objects.  We
present the photometric light curves from 2014 to 2017 obtained by the
Steward Observatory's Bok telescope and the CFHT telescope with
MegaCam.  We provide details on the data acquisition and processing of the data
from each telescope, the difference imaging photometry used to produce the
light curves, and the calculation of a variability index to quantify
each AGN's variability.  We find that the Welch-Stetson J-index
provides a useful characterization of AGN variability and can be used
to select AGNs for further study.
\end{abstract}


\section{Introduction}
\indent
The variability of active galactic nuclei (AGN; hereafter synonymously
used with ``quasar") is one of the most powerful tools available to
explore the physical processes in these objects and to determine their
role in the evolution of galaxies (e.g., \citealt{Ulrich97,
  Vandenberk04, Macleod10}). 
Various aspects of quasar variability touch on a wide range of
physical processes \citep{Kozlowski:2010}, and studies span a wide
breadth of topics, ranging from continuum variability and the
exploration of quasar accretion processes (e.g., \citealt{Krolik91,
  Clavel91, Czerny99, Kelly11}), to emission-line variability and the
broad-line region (e.g., \citealt{Blandford82, Peterson93, Kaspi:2000,
  Sun15}), to the study of so-called ``changing-look" AGN (e.g.,
\citealt{Tohline76, Cohen86, Clavel89, Denney14, Shappee:2014,Lamassa15,
  Macleod16}), and to studies of absorption line variability (e.g.,
\citealt{Smith88, Turnshek88, Barlow93, Lundgren07,Filiz:2013}). 
AGN variability/monitoring programs provide important
insights into the nature of the inner regions of quasars on a variety
of scales, leading to better understanding of their role in the
evolution of AGN and galaxies. 
There are many time-series datasets that focus on AGN over a variety of
scales.  Most spectroscopic monitoring projects, by necessity,
have focused either on small numbers of AGN at high cadence or large
numbers of AGN at very low cadence. 
While there are science goals that can be accomplished by both
types of datasets, statistical interpretations are limited by the small
sample sizes in the first case, and detailed investigations are limited by the small numbers
of epochs in the second case. 
Recently, however, large surveys that use multi-object spectrographs
have begun to target statistically significant samples of
representative quasar populations for much higher cadence,
long-duration studies (e.g., \citealt{King:2015}). 

The Sloan Digital Sky Survey Reverberation Mapping (SDSS-RM) project
\citep{Shen:2015} is an ongoing program that monitors 849 
broad-line quasars/AGNs. 
The project is based on a spectroscopic campaign using the Baryon Oscillation Spectroscopic
Survey (BOSS) spectrographs \citep{Smee:2013} that began at the end of the SDSS-III survey and
continues through the SDSS-IV survey \citep{Blanton:2017}.  As of summer 2019, the program has accumulated 6 years of SDSS spectroscopic data.
To complement the spectroscopic
data, the SDSS-RM field has been photometrically monitored at the Canada-France-Hawaii (CFHT) and Steward Observatory Bok
telescopes.  The SDSS-RM field was observed in the SDSS {\it g} and {\it i} bandpasses with the CFHT
from 2014 to 2016, and with the Bok from 2014 through 2017. 
In addition, the field was monitored photometrically as a part of the
Pan-STARRS1 MD07 program during 2010-2013, which extends the total time
baseline to 9 years. 
This program will continue to be observed through 2020, and monitoring
of this field will continue as a part of the SDSS-V Black Hole Mapper project (BHM; \citealt{Kollmeier17}). 

Reverberation mapping (RM; \citealt{Blandford82, Peterson2004}) is the
primary objective of the SDSS-RM program.  RM uses the time
delays between the variability of emission from different
regions within the AGN. 
For example, optical continuum variations are
echoed by emission lines from the broad-line region (BLR), but
with a time delay corresponding to the time it takes for light
to travel to the BLR from the continuum-emitting region. 
This information yields a characteristic radius of the BLR gas, which
can be combined with an orbital velocity estimate from the width of
the broad emission lines to estimate the black hole mass.

The main goals of SDSS-RM are to obtain reverberation mapping BLR
time-lag measurements and to estimate black hole masses for hundreds of quasars.
Technically, these goals require only spectroscopic monitoring. 
However, the cadence of the SDSS-RM spectroscopy, while higher than
any previous study of such a large sample, is by itself inadequate. 
Supplemental photometric monitoring of the SDSS-RM field is critical,
as the additional photometry can better sample
the continuum light curves, led by
improved sensitivity to shorter time lags and 
improved accuracy for longer lags.  
In addition, the photometry can be used to improve the spectrophotometric calibration of the spectroscopy. 
Simulations and real data anlayses thus far indicate that
the additional photometry will allow us to nearly double the number of
lag determinations compared to only using the spectroscopic data
\citep{Shen:2015, Shen16b, Grier:2017}.

The SDSS-RM program has reported 
$\sim100$ additional broad-line lag measurements (\citealt{Shen16a};
\citealt{Grier:2017}; \citealt{Grier:2019}).
In addition, the SDSS-RM collaboration has produced a wide
variety of results using the first 1--4 years of data,
including studies of broad absorption-line variability, host-galaxy
properties, emission line velocity shifts, and
quasar variability (\citealt{Shen15b, Grier15, Sun15, Matsuoka15,
  Shen16a, Shen16b, Denney16a, Denney16b, Li17, Grier:2017, Sun18,
  Yue18, Hemler19, Wang:2019}). 
However, there will be many other scientific applications for a
dataset containing more than 70 spectroscopic epochs over a period of
several years with more than 100 epochs of accompanying photometric
observations of a sample of 849 quasars that span a wide range in redshift $(0.1 < z < 4.5)$,
and were chosen to span a wide range quasar properties (see \citealt{Shen19}).  

In this work, we present the details of the photometric data from the
CFHT and Bok telescopes obtained during the first four years
of the SDSS-RM program. 
Details of the telescopes and the strategy for data collection at both
facilities are described in Section 2. 
The datasets were individually pre-processed by their respective
facilities' pipelines (see the subsections in Section 2). 
The light curves were produced using the ISIS difference imaging
packages \citep{Alard:2000} described in Section 3.
We present in Section 4 the light curves for 
all 849 SDSS-RM targets and an additional 365 AGNs that were also
monitored as a part of the program by the Bok telescope. 
In Section 5 we calculate a variability index to quantify
the variability of each object and to identify which objects have a
highly variable continuum.  We conclude this
paper in Section 6 with a summary of our work and provide a synopsis
of future studies.

\section{Data Collection}

\subsection{CFHT Observations}
The CFHT/MegaCam (Aune et al.\ 2003) 
is a wide-field optical imager with a pixel scale of $0.\arcsec187$.
Its field of view (FoV)
was expanded from 1 deg$^2$ (36 CCDs) to 1.1 deg$^2$ (40 CCDs) in 2015. 
All MegaCam observations were made in Queued Service Observations (QSO) mode,
and a log of the observations is presented in Table \ref{cfhtlog}.
Each semester's observations at the CFHT are discussed in the
appropriate section below.  Raw and calibrated CFHT/MegaCam images
can be retrieved from the Canadian Astronomy Data
Centre using the proposal IDs given below.\footnote{\href{http://www.cadc-ccda.hia-iha.nrc-cnrc.gc.ca/en/cfht/}{http://www.cadc-ccda.hia-iha.nrc-cnrc.gc.ca/en/cfht/}}

\subsubsection{CFHT pointings layout}
\begin{figure}
\includegraphics[scale=0.45]{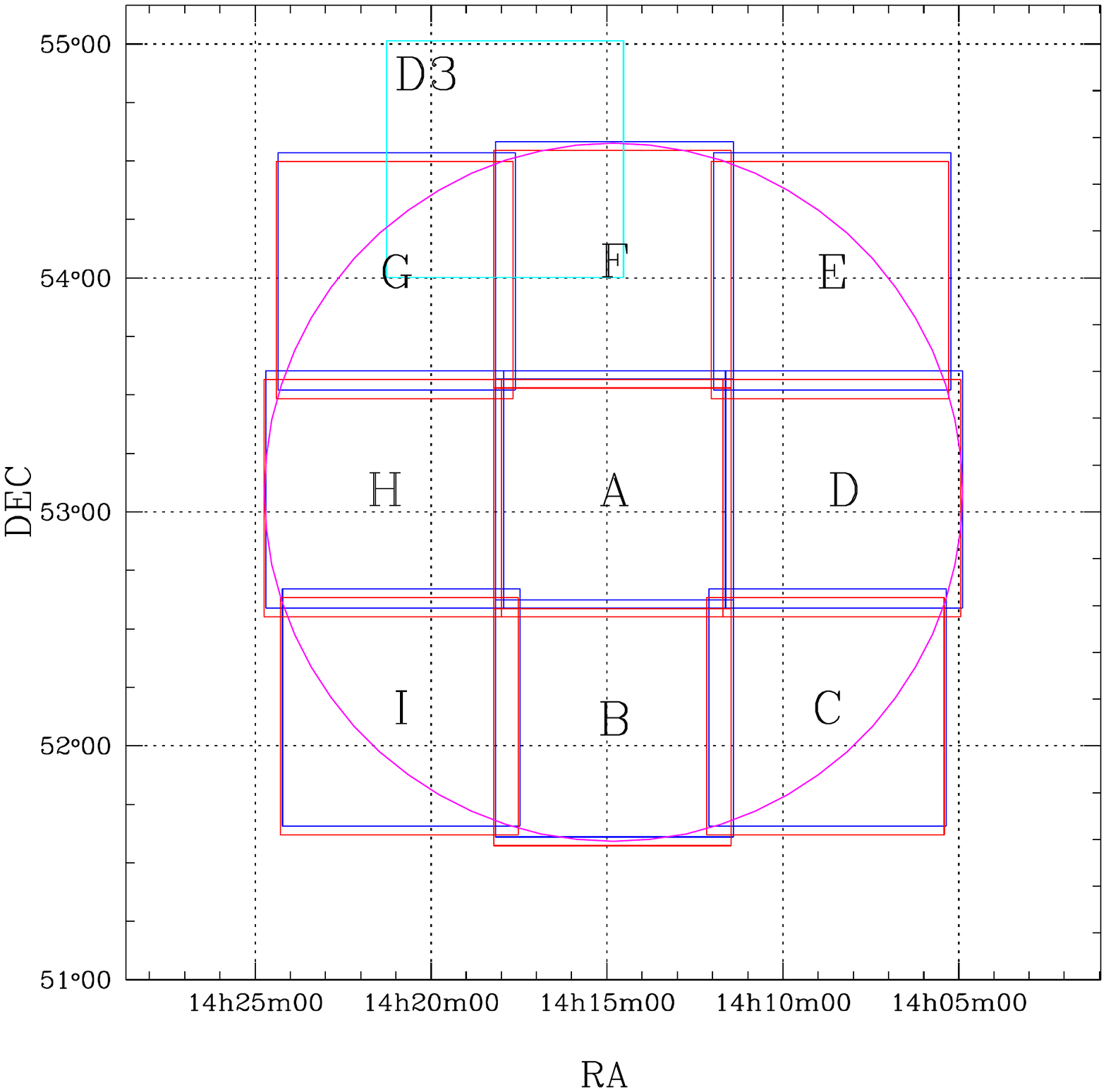}\includegraphics[scale=0.45]{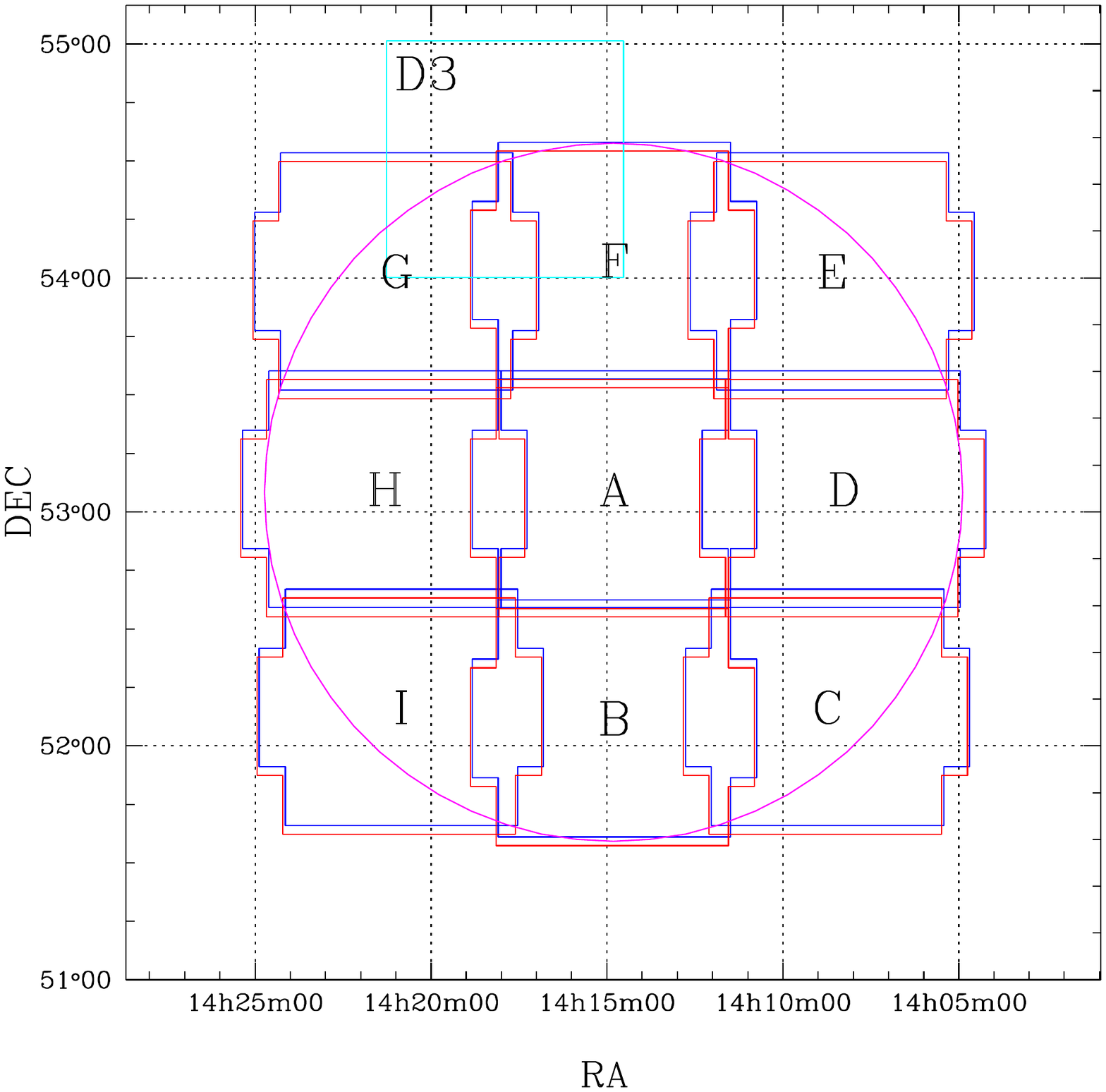}
\caption{RM field pointing layout.  The magenta circle is the
  SDSS-BOSS spectrograph FoV.  The SDSS-RM field is contained
  within the CFHTLS W3 (wide) field, and partially overlaps the CFHTLS
  D3 (deep) field, indicated by the cyan square labeled ``D3''.  The
letters denote the positions of each of the 9 pointings, and the order in which they are
observed.  The blue and red outlines are the two dithered
exposures  at each pointing.  Each pointing overlaps the next by $\geq
16\farcs5$ in RA and $\geq 88\farcs5$ in DEC.  The left panel shows the imaged
areas of the FoV
for semesters 2014A and 2015A, when 36 CCDs were unvignetted.  The
right panel shows imaged areas of the FoV for semesters 2016A and
2017A, when 40 CCDs were unvignetted.  Both are approximate, in
that they do not account for the small change in the extent RA of
the MegaCam field of view over the range in DEC shown.
\label{fig:f_pbh12}}
\end{figure}

The SDSS-RM field overlaps with the W3 wide-field and the D3
deep-field of the Canada-France-Hawaii Telescope Legacy Survey
\citep{Semboloni:2006}\footnote{See
  \href{http://www.cfht.hawaii.edu/Science/CFHTLS/T0007/T0007-docsu10.html}{http://www.cfht.hawaii.edu/Science/CFHTLS/T0007/T0007-docsu10.html}}.
 To cover the footprint of the $7^{\circ}$ SDSS-RM field,
we used a total of nine slightly overlapping pointings labeled A through I
with two dither positions at each pointing to fill in the CCD gaps.  
The sky coordinates of the centers of the 18 different dither positions
(labeled A1 through I2) are given in Table \ref{table:t_pbh1}.
The layout of the fields is shown in Figure \ref{fig:f_pbh12}a for 2014A-2015A (36 unvignetted CCDs)
and in Figure \ref{fig:f_pbh12}b for 2016A-2017A (40 unvignetted
CCDs).  Two exposures were taken at each dither position.
The first was a fixed 30 (40) s short exposure in $g$ ($i$) for bright RM targets.
The second had an exposure time determined dynamically
to achieve a desired point-source S/N of 25 at $i = 22$ and $g = 22.25$,
given the seeing and transparency at the time of the observation.
At CFHT, this procedure is known as SNR-QSO mode. The average dynamic exposure
lengths in 2014A were 55 s in $g$ and 77 s in $i$. 

\subsubsection{Observing Semester 2014A (Proposal 14AC02)}
A total of 81.4 hours was awarded from Canada, France and China, and 52.2 hours of data were obtained.
The observations were made
using the CFHT filters g.MP9401 and i.MP9702 \footnote{See \href{http://www.cfht.hawaii.edu/Instruments/Filters/megaprime.html}{http://www.cfht.hawaii.edu/Instruments/Filters/megaprime.html}}. 
In each MegaCam run, the $g$ and $i$ observations were executed 
in Observation Groups (OGs).  Completion of the $g$ observation
group triggered, via a relational execution link (REEL), an
$i$ observation group to be executed within 12 hours (24 hours in
Dec/Jan and Jun/Jul). On most occasions, the $g$ and
$i$ OGs were executed on the same night. If no $i$ epoch was
obtained within 24 hours of a $g$ epoch, the $i$ epoch was skipped in
favor of a new $g$ epoch two nights later.  Under poor weather
conditions, the requested cadence increased to nightly near the
end of a MegaCam run to try to obtain a minimum of 4 epochs.

\subsubsection{Observing Semester 2015A (Proposal 15AS02)}
A total of 20.4 hours was awarded from Canada, France, and China,
and 15.5 hours of data were obtained.
At the beginning of 2015A, Megacam was upgraded from 36 to 40 CCDs.
New filters that allowed unvignetted observations of the entire FoV  
were also installed.  For our observations, however, we wished to 
avoid any changes in photometry due to bandpass differences;
thus, we continued to use the same filters as in 2014A.
In this semester, we tried to obtain two $g$-band epochs 
per MegaCam run, separated by 4 to 10 days.  Additional $g$-band
epochs could be added to a run if the time allocated for our
observations in previous MegaCam runs had gone unused.
An $i$-band epoch was requested to occur either
within 15 hours after a $g$-band epoch or not at all.

\subsubsection{Observing Semester 2016A (Proposal 16AC12)}
 A total of 20.4 hours was awarded from Canada, France and China,
and 15.8 hours of data were obtained.
Beginning with semester 2016A, we used the new filters
(CFHT filters g.MP9402 and i.MP9703 \footnote{See \href{http://www.cfht.hawaii.edu/Instruments/Filters/megaprime.html}{http://www.cfht.hawaii.edu/Instruments/Filters/megaprime.html}})
for an unvignetted view of the upgraded focal plane with 40 CCDs.
However, we have not yet processed the photometric measurements of the target
quasars using images from the 4 new CCDs.
We used the same scheduling pattern as in semester 2015A.

\subsubsection{Observing Semester 2017A (Proposals 17AC12 + 17AF13)}
 A total of 16 hours was awarded from Canada and France,
and 12 hours of data were obtained.
Due to a smaller time allocation in 2017A,
we chose to request only $g$-band observations.
After the entire field was observed, at least 72 hours 
were required to elapse before it could be observed again.
Results incorporating these data will be presented at a future date.

\subsection{Bok Observations}
Monitoring of the SDSS-RM field with Steward Observatory's 2.3m Bok telescope commenced in December
2013 and is ongoing. 
The SDSS-RM observations were allocated time through the regular proposal process
and observing nights were often shared with other programs to maximize time
coverage and observing efficiency. SDSS-RM data are collected with the
90Prime instrument \citep{Williams:2004}, a $\sim1^\circ$ field of view.
As with the CFHT observations, Bok data are collected in the
SDSS $g$ and $i$ bandpasses. To complement the CFHT dark time observations and
provide the continuous photometric coverage, the Bok observations were typically
obtained during bright lunar phases, including full Moon. The sky background is
thus quite high, which degrades the overall sensitivity. Each Bok pointing
has a design exposure time of 300s, but this time was sometimes divided into multiple
exposures to keep the bright sky background well below
saturation limits.

\begin{figure*}[!h]
 \plotone{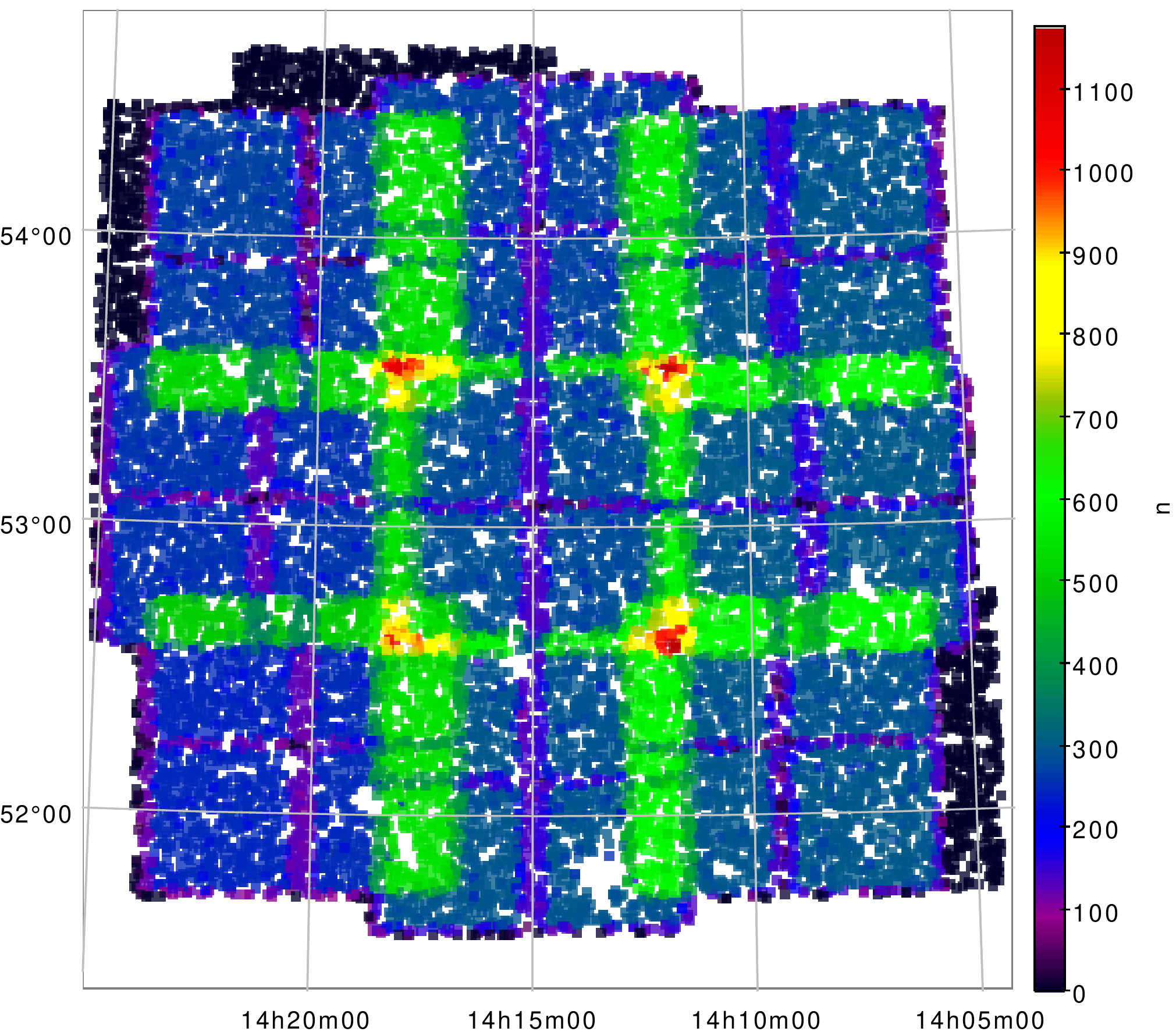}
 \caption{Number of observations of individual standard stars in the
   Bok observations of the SDSS-RM field,
 including data obtained during 2014-2016. The patterns due to the pointing sequence are
 clearly visible: dark blue regions are observed only once during a pointing 
 sequence, and light blue regions are observed twice due to the overlap in dithered
 frames at each of the 9 primary pointing centers. Green regions are overlaps
 between the 9 pointing centers, resulting in twice as many observations per
 star. Finally, the yellow/red regions near the center are where 4 or more
 pointings overlap. Holes are regions around bright stars. Over the three year 
 period, individual stars received anywhere from $\sim$200 to $\ga$1000
 photometric measurements.
 \label{fig:coverage}}
\end{figure*}

The 90Prime FoV covers the SDSS-RM field in 9 total pointings. To better cover
CCD gaps and to compensate the high backgrounds, these pointings are divided into two
exposures (typically 150s, as above) that are dithered by a few arcminutes,
leading to a total of 18 pointings per epoch. Usually only one bandpass is used per
night and filter changes occur on adjacent nights. An exception is the three 
nights around full Moon, where only the $i$ band is used due to the high 
background at shorter wavelengths. Collecting
all 18 pointings takes roughly two hours with overhead for focus and pointing
alignment. The pointing sequence is repeated through the night as time
permits. Given the overlap between pointings as well as the repeated passes,
an individual quasar receives anywhere from one to $\sim$10 observations
in a single night, allowing for analyses of short-timescale variability. The
observation pattern is presented in Figure~\ref{fig:coverage}, demonstrating
that regions within the SDSS-RM field received between $\sim200$ and $\ga$1000
observations over the three year period beginning with the 2014 campaign, obtained
from over 5000 individual exposures.  Table~\ref{bokobslog} provides
the log of the Bok observations over the four years.  

\subsubsection{Bok Data Reductions}
Bok 90Prime data were processed using standard optical image processing techniques 
implemented in a python-based pipeline developed by our group 
\footnote{https://github.com/imcgreer/idmrm and https://github.com/legacysurvey/rapala}.
The pipeline organizes data by observing night and automatically
identifies the calibration frames taken for each night. Bias and dome flat correction frames are
produced by a stacked median of $\sim$10 calibration frames. The science frames
are sorted by filter and associated with the calibration frames that
are nearest in time.  The overscan region of each science image is collapsed to one dimension using a
median, followed by a 17-pixel wide median filter to smooth the overscan pattern.
The resulting vector is subtracted from the image. The two-dimensional bias
frame is then subtracted, and the dome flat is applied to the images.

The four 90Prime CCDs each have four readout channels with independent amplifiers.
The next step is to combine the 16 image channels into a single, 4-CCD mosaic with
the necessary gain corrections. We found that there are small gain variations over
the course of a night and that tracking these variations is necessary to obtain
a uniform sky background. To remove sensitivity to illumination gradients, we 
selected small windows of pixels along the borders of adjacent images, then fit
for the multiplicative factor needed to balance the sky levels of the associated
amplifiers. Once the four amplifiers in each CCD were balanced to a common level,
we used a similar procedure to fit the sky levels near the center of the field
in order to match the CCD sky levels.

The dome screen illumination pattern is quite different than the night sky illumination, so the science images 
from each night are combined into a filter specific illumination correction. This 
correction is obtained by scaling each image to a common sky background level
and stacking with a clipped mean. The image is binned by a factor of 8 along
each axis using a median to remove small objects (stars and galaxies). A smooth
2D spline is then iteratively fit to this rebinned image, removing outlier pixels.
The final result is applied to the science images and results in a highly uniform
sky background.

For the $i$-band images, we also create a fringe pattern image. The
median sky level is removed from each image, and the images are stacked to produce
the fringe master. The sky-subtracted images are divided by the fringe master
to determine the scaling of the pattern that matches the fringe amplitudes in each
frame. Then, the scaled fringe master image is subtracted from the original science frame (prior to removing the median
sky level).

A final, high signal-to-noise ratio sky flat is constructed from the science
frames by median combining the images scaled by the sky background after masking
bright objects identified using Source Extractor \citep{Bertin:1996}. This sky flat is usually composed of
images from a single observing run, unless there are too few ($\la$100), in
which case they are combined over multiple runs. Although the corrections in the
sky flats are small (few per cent), there are notable patterns remaining after
the previous processing steps that the sky flat is able to correct.

The final detrending step is to remove a low-level variable gradient
in the sky by fitting a linear two-dimensional spline to the sky background of each image after
masking objects. Only the gradient is subtracted -- the mean sky level of each
image is preserved.  Object catalogs are generated using SExtractor
and then Scamp \citep{Bertin:2006} is used for astrometric registration.
A catalog of astrometric references for the field was initially constructed
from SDSS, and later updated to use Gaia DR1 positions \citep{Lindegren:2016}. The resulting RMS in star
positions is $\sim$0\farcs05.

The pipeline utilizes multiprocessing to run processing steps in parallel. On
a typical multi-core desktop machine, the entire set of $>$5000 images can be
processed in roughly one day.

\subsubsection{Photometric Analysis}
Although the primary source of photometry for SDSS-RM is the difference imaging
catalogs, we also performed aperture photometry on the Bok images. Locations
of both reference stars from SDSS and the RM quasar targets were identified
using the WCS solutions from the processing pipeline. A set of apertures from
0\farcs9 to 10\farcs1 were placed at each target location and counts within
the apertures were obtained with \texttt{sep}.\footnote{https://github.com/kbarbary/sep}
We primarily focus on the 2\farcs3 and 6\farcs8 apertures, which
provide roughly optimal and total light apertures for typical Bok seeing.

We construct light curves for all targets. The reference stars are treated
as constant sources and are a valuable cross-check on processing errors or
poor quality imaging that leads to photometric outliers. For each reference star,
we determine a median magnitude and RMS scatter, and then flag measurements that are $>5\sigma$
outliers. If a single image contains a large number of outliers, we examine
it by eye. Most of these cases are images affected by strong scattered light from the
bright Moon during the Bok observations, and they are removed from our analysis.
Once the data set is cleaned of bad epochs,  we determine photometric zero points and color
terms relative to the SDSS $g$ and $i$ band photometry of the field using bright
($g<17,i<19$ mag) stars.

After removing outliers, we find that the internal accuracy of the Bok photometry
reaches $\sim10$ mmag for bright stars (17 mag), and $\sim30$ mmag at $g$=19. This
suggests we have met our goal for photometric accuracy and that the Bok photometry
will be suitable for tracking quasar variability.

\section{Difference Imaging}

For the analysis of the data sets collected at the CFHT and Bok
telescopes, we used the  ISIS difference imaging package \citep{Alard:1999,
  Alard:2000}.\footnote{The software can be downloaded from
  http://www2.iap.fr/users/alard/package.html}  
To produce the differential flux light
curves, the reduced data were parsed into separate FITS files,
separated by field, CCD, and bandpass.  The CFHT data had 9
fields while the Bok data had 18.  The CFHT
and Bok data sets were processed with ISIS separately.

The images for each subfield are first interpolated to a common
frame.  Next, a reference image is constructed using the images
with the best seeing, lower sky backgrounds, and high transparencies.
For each individual image, the reference image is convolved to match
its PSF structure, and the individual image is scaled and subtracted
from that convolved reference image to create a difference image on
the flux scale of the reference image.  Negative values in a
difference image indicates that a source is brighter than on
the reference image.
Details of the convolution matrix and their applications are described in
\citet{Alard:1998} and \citet{Alard:2000}. 

The RM targets and a selection of standard stars were identified on
the reference image, and PSF photometry was performed at those positions.
Using the midpoint of the image observation (in terms of Modified
Julian Date) as our time index, light curves were constructed with the differential
flux values from ISIS.  In this paper, we present the differential
flux light curves for each SDSS-RM quasar from each field in which
it appears on that field's reference image.  \citet{Grier:2017} describe how for their analysis,
they merged these differential light curves from the different
datasets and corrected the uncertainties reported by ISIS using the
standard star observations.  

\subsection{CFHT Processing}
For the CFHT data spanning 2014 to 2016, the raw data were detrended
with the Elixir pipeline.  With the detrended data,
we separated the multi-extension FITS files into the individual CCD
images (36 for the 2014 data set and 40 for the subsequent years).
Processing was systematically performed per field, per pointing (1 or
2), per filter ($g-$ or $i-$ band), and per CCD.  Source Extractor
\citep{Bertin:1996} was used to identify the best
subset of the images to create the reference image, based on the sky
background and the median FWHM.  The number of images
selected to construct the reference image ranged between 4 and 10.
For some subfields, a bright foreground star would dominate and
inflate the sky background.  In such problematic subfields, fewer images
could be included in the construction of a robust reference image.

Each CCD has a small field of view and typically contained 2 to 8
AGNs.  We selected $\sim 100$ reference stars in each subfield along
with the AGNs.  We used the standard stars' light curves to rescale the
uncertainties and correct for any offsets between the years (e.g., due
to the filter change - see Section 2.1).

Once the image quality had been assessed and all the targets
identified in the fields, we began the ISIS
difference imaging processing by doing an interpolation of the images,
reference image creation, image subtraction of the convolved images to
the reference image, and finally PSF photometry of the residual
images.

\subsection{Bok Processing}
The dataset from the Bok $90''$ telescope spanned the observational
season of the SDSS-RM field from 2014 to 2017.  The field was observed
in mostly bright time, which allowed coverage of the variability during
the times when the field was not observed by SDSS or CFHT.  The bulk
of the data were primarily taken during 2014, and it is this dataset
that serves as the baseline for the difference imaging processing of
the subsequent years of Bok data. 

The Bok data from 2014 were processed with ISIS mostly using the same
procedures as for the CFHT 2014 data, but with a few changes.  One
extra step in the preparatory work done to the Bok images was to add back
in the sky background value, as this information was removed from the
reduced images from the Bok pipeline.  Without this sky background
information, ISIS had difficulty producing a useful reference image as
well as convolving the rest of the images because ISIS assumes the sky
is included.  This particular step was done only for the 2014
dataset. In the 2015-2017 data reduction, the sky background was
retained.  Next, the image with the best FWHM and lowest sky background was
identified, and the other images were interpolated to its reference
frame.  This reference image (which was based on primarily 2014
images) was the basis for the later, incorporated Bok data sets.  Since the Bok
fields had a larger field of view, more reference stars were available
and processed per field.  The standard stars were used to 
rescale the uncertainties in the flux measurements of the SDSS-RM AGNs (see \citealt{Grier:2017}).

In addition to the original 849 RM targets from \citet{Shen:2015}, 365
objects were also identified as possible AGN in the Bok fields.  These extra 365
targets currently lack spectroscopic data and only have Bok time-series
photometry spanning the four years.  These AGN are in the CFHT data,
but at the time of the ISIS processing, the information for these
extra objects were not available, and thus they were not processed.

\section{RM target light curves}
For the photometric light curves, we include with this paper the ISIS
output from its {\it phot} module of the ISIS software.  The output photometry
files are provided as online data and are archived as a
multi-extension FITS file.  The columns in the output photometry files are as follows:  

\begin{description}
\item[Column 1] Midpoint of the observation in Modified Julian
Date (days).  
\item[Column 2 and 4] Differential fluxes 
\item[Column 3 and 5] Respective uncertainties in the
  differential flux measurement of Column 2 and 4 
\item[Column 6 and 7]  Sky background measurements 
\item[Column 8] The local average sky value over Nsky pixels. 
\item[Column 9] The local median sky value over Nsky pixels.
\item[Column 10] Nsky pixels used in Column 8 and 9.
\item[Column 11] The aperture flux measured over N pixels.
\item[Column 12] The number of pixels, N
\end{description}

The fluxes are in units of counts but are all on the flux scale of
their reference image.  Two differential flux values are calculated, which are based
on two different PSF weighting schemes for extracting the fluxes. Any significant difference between these fluxes is an
indication of subtraction problems usually due to a nearby bright
source, clouds, or other systematic problems.  ISIS itself assumes
clean subtractions, in which case the ``sky'' on a subtracted image is
consistent with zero.  However, clouds can produce a local, variable ``sky'' in the
subtracted images, and we use a modified code which attempts to
correct for a local sky.
For our analysis we used the flux and
uncertainty values of Columns 2 and 3.  The actual photometric noise
may be larger, by up to a factor of 2, compared to what is reported in Columns 3 and 5. 
The crude aperture flux value in Column 11 can be used as a sanity 
check against the flux values determined in Columns 2 and 4.  
In the ISIS convention, a negative flux on the subtracted image means
the source is brighter than on the reference image.  To convert the differential
flux to a magnitude, the flux of the object from the reference image
must be subtracted to the {\it differential flux}.  These differential flux
values are all with respect to the reference images used in the
processing.  We intend to provide the reference flux values of the
QSOs in a later paper after incorporating additional datasets that are
being collected now.  

Since the field of view of both the CFHT and Bok data sets are
relatively large, we can see variable cloud cover in the frames.  The
differential flux scalings are affected by this cloud cover, and by
the halos of bright, foreground stars.  In both cases, the sky
background values can be non-zero, and can be checked with the values in
Column 6 and 7 of the ISIS output files.
A number of targets found close to the edge of a CCD may have fewer
epochs due to the object occasionally falling off the edge of the
image.  Due to the dithering, the edges of the reference
image are some 20 pixels inward from the nominal chip edge.  The
software would discard flux measurements of targets from these edge
regions due to problems such as poor PSF profiles.
In some cases, these edge targets have fluxes measured from an
adjacent subfield, so the light curves can still be constructed.

\section{Measuring AGN Variations}
The photometric dataset we present here provides an excellent
opportunity to apply a variability index, which can be used to
select interesting targets for further study.  We characterized the
variability with the Welch-Stetson J-index, which was originally developed for
identifying pulsating variable stars \citep{Welch:1993, Stetson:1996}.
We calculated the J-index using both the Bok
and CFHT datasets; however, the Bok dataset has the advantage of
having more epochs, which yielded a more robust variability index.
The description of the Welch-Stetson J-index construction is provided in the appendix.
For a more thorough
description of this analysis tool, we direct the reader to \citet{Stetson:1996}.

\subsection{Application of the variability index}

\begin{figure}
\begin{center}
\includegraphics[scale=0.7,angle=0]{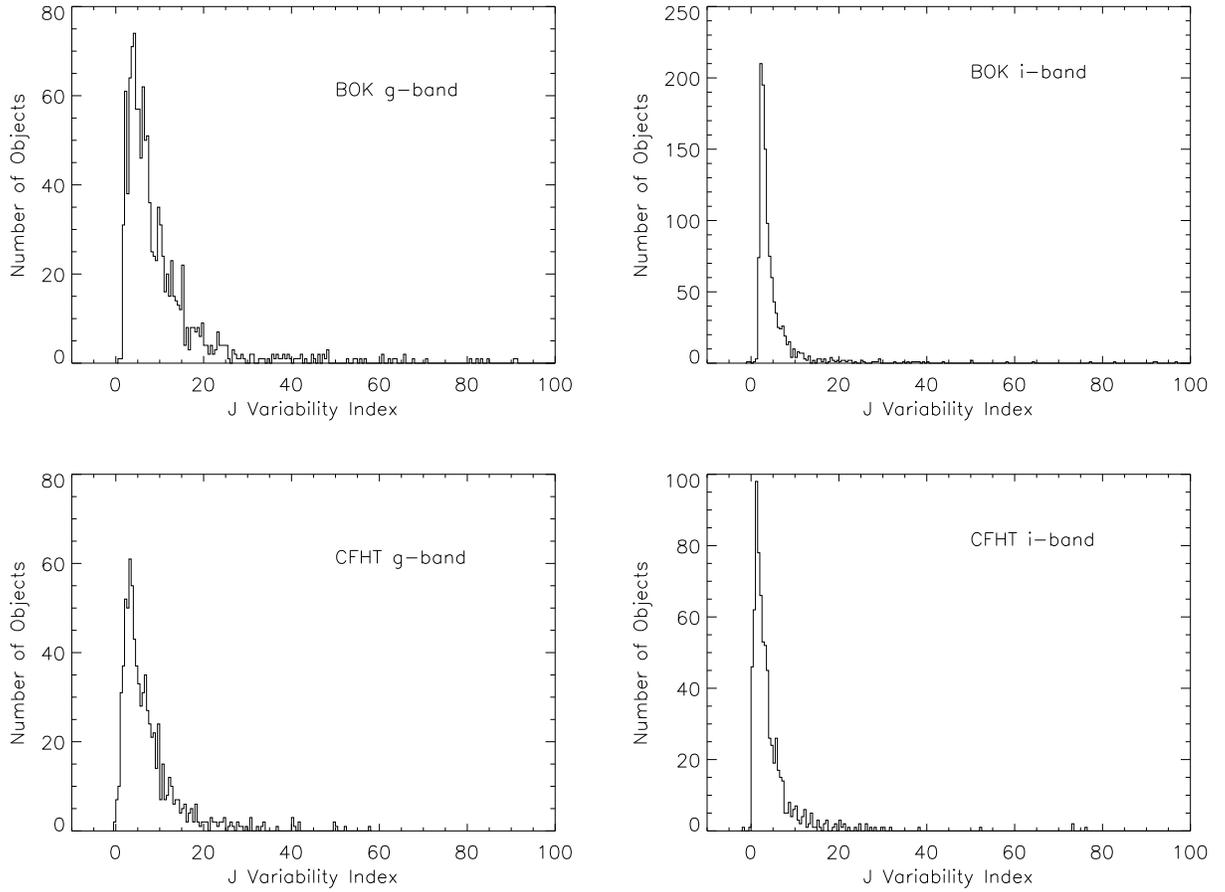}
\caption{Distribution of the J variability indices for the SDSS-RM
  AGNs.  Upper plots are for the J variability indices
  calculated from the Bok datasets.  Lower plots are
  for the J indices calculated from the CFHT datasets.  The left plots are
for $g$-band data and the right plots are for $i$-band data.}
\end{center}
\label{fig:histograms}
\end{figure}

We calculated the variability index for all 849 SDSS-RM objects
using both the Bok and CFHT $g$ and $i$ data.  We also calculated
the Welch-Stetson J index for the additional 365 AGN \citep{Shen19} with the Bok data.  These
additional AGN were not processed in the CFHT datasets, since at the
time of the differential flux processing, the positions of those
objects were not available.  Subsequent ISIS processing of the CFHT
dataset is planned to include these additional AGNs.  The Bok dataset has
more epochs and a longer time baseline. Thus, the Bok variability
index should be more accurate and better reflect the variable nature of the
AGN.  We find that the J-index values calculated from the CFHT dataset
are consistent with the values based on the Bok datasets.  We
provide the J variability index for each SDSS-RM AGN in Table
\ref{varidx}. In Figure \ref{fig:histograms}, we present histograms of
the J-index with respect to filter and telescope.  Our calculation of
the J-index took into account of the case where an AGN was observed over
different subfields due to the dithering of the field.  Separate
J-indices were determined for the table, and we averaged those values.

\citet{Grier:2017} examined 220 AGNs of the SDSS-RM sample 
for RM lags between the continuum and the $H\beta$ line.  To select
AGN that were variable, they used a measurement of the RMS scatter about the mean
to quantify the variability.  Their method differs from the formulation of the
Welch-Stetson index; however, a visual inspection of the light curves
indicates that both measurements track AGN variability well.

We note that there were several differences in the treatment of the
light curves before these variability measurements were made.  For
example, \citet{Grier:2017} merged the light curves from the different
datasets and corrected the flux uncertainties from the original ISIS
fluxes.  In our work, we adopted the differential flux values without any
correction.  Thus, a direct comparison between our J-index and their
RMS variability measure is not feasible.

\begin{figure}
\begin{center}
\includegraphics[scale=0.6,angle=0]{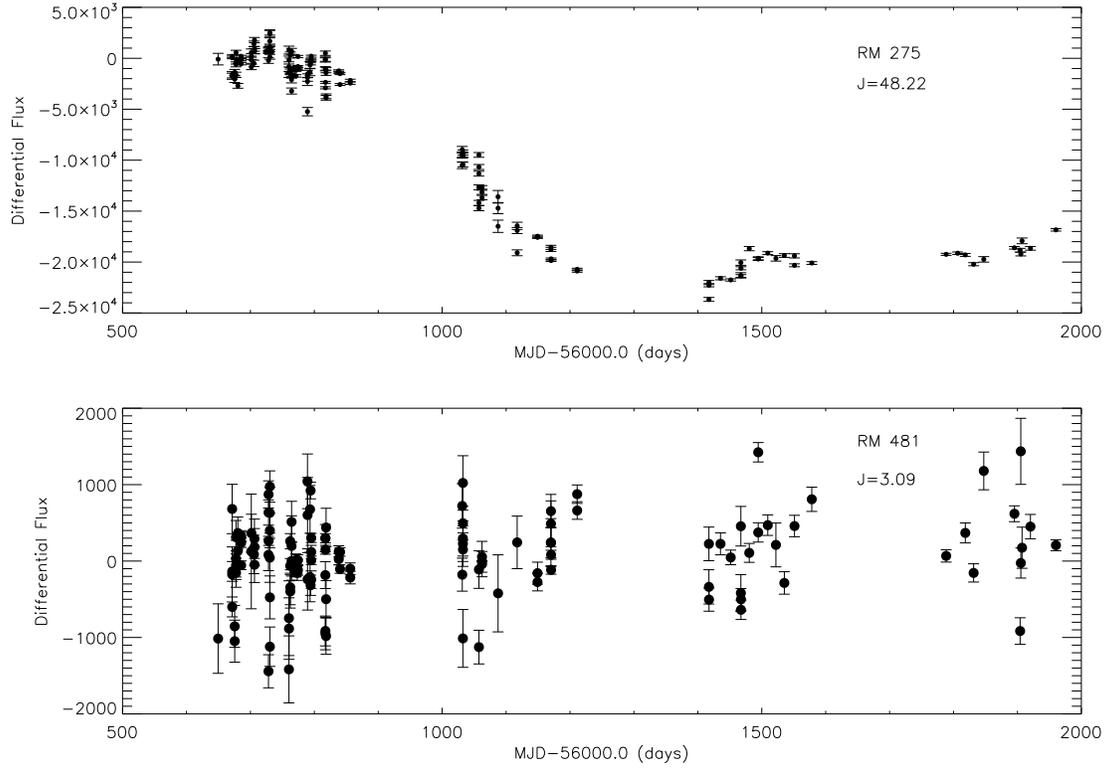}
\caption{(Upper)  Light curve of RM 275, which has a large, long-term variation. RM 275 has a calculated
variability index of $J=48.22$.  RM 275 has a $g$-band magnitude of
20.73.  The photometric error bars are negligible with
respect to the range in flux presented here. 
In the ISIS convention, a negative differential flux means the source
is mostly brighter than on the reference image.
(Lower) Photometric light curve of RM 481, which appears to
  have a relatively small variation between 2014 and 2017.  This AGN
  has a $g$-band magnitude of 20.74 and a variability index of $J=3.09$.}
\end{center}
\label{fig:ltcurve}
\end{figure}

Determining what level of variability measurement constitutes 
``significant'' variability is somewhat subjective.  Our visual
inspection of the 220 AGN sample suggests that the J-index is well
correlated with the actual variability (both in the short and
long-term).  In Figure \ref{fig:ltcurve}, the light curves of two AGNs
with similar $g$-magnitudes show how the J-index can be used to pick
out highly variable objects.  RM 275 has a larger J-index value
compared to RM 481, and their light curves exhibit their levels of
variation reflecting their different variability indices.  We provide here the J variability index as a guide for
readers to select interesting objects for future projects.  An example
of a large variability index indicating activity in an AGN is RM 17,
which is described as a hypervariable quasar by \citet{Dexter:2019}.

\begin{figure}
\begin{center}
\includegraphics[scale=0.7,angle=180]{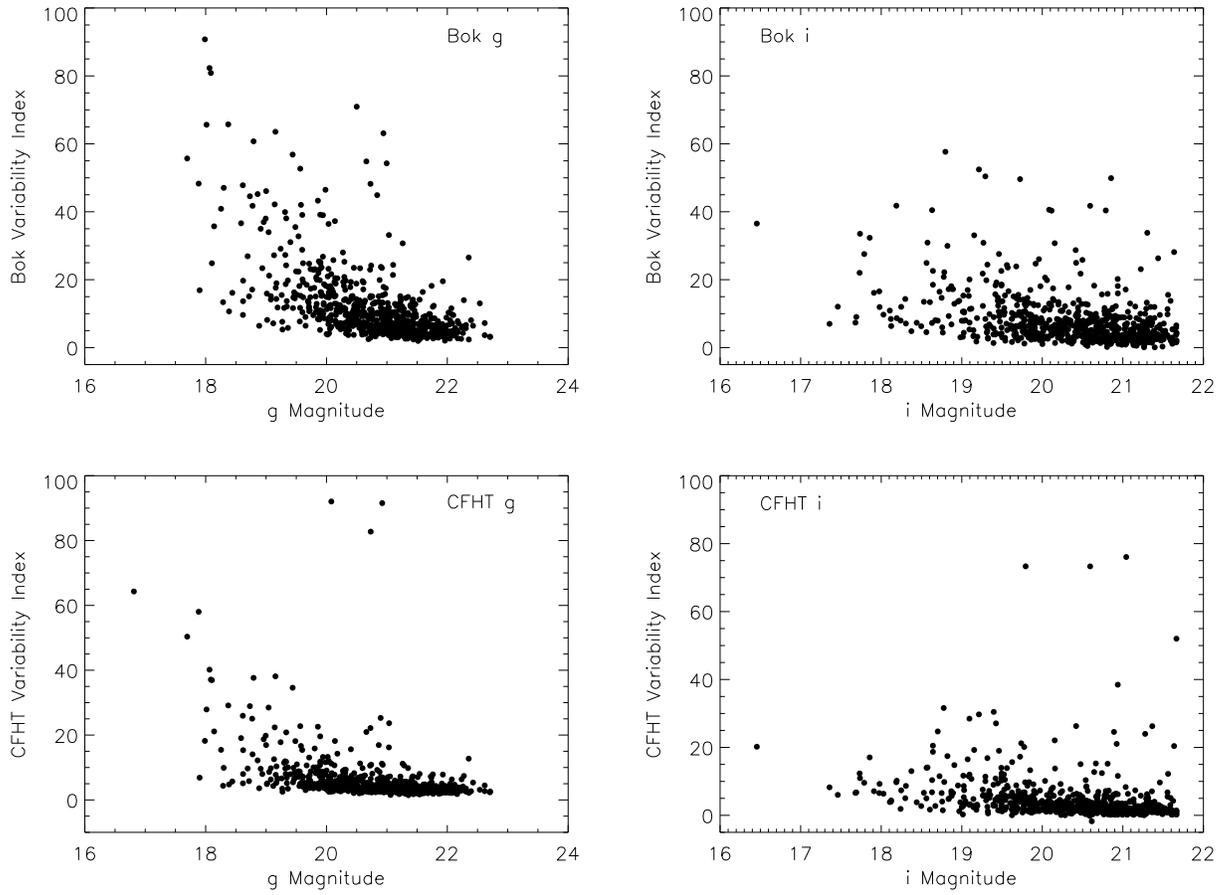}
\caption{(Upper) Bok $g-$ and $i-$band variability indices as a
  function of magnitude for the SDSS-RM AGNs.  (Lower)
  CFHT $g-$ and $i-$band J-indices as a function of magnitude.}
\end{center}
\label{fig:jidxmag}
\end{figure}

Figure \ref{fig:jidxmag} presents the variability index with respect to the $g$ and $i$
magnitude of the SDSS-RM objects.  We show in this Figure the
distribution of the indices from the Bok datasets and the CFHT datasets.  At brighter
magnitudes, larger variability indices are seen, but at fainter
magnitudes, we find fewer objects with such a level of variability.
This result could be due to an observational bias in the sample, but
in general, objects with large variations at brighter magnitudes are
more readily identified in a sample.

\begin{figure}
\begin{center}
\includegraphics[scale=0.7, angle=0]{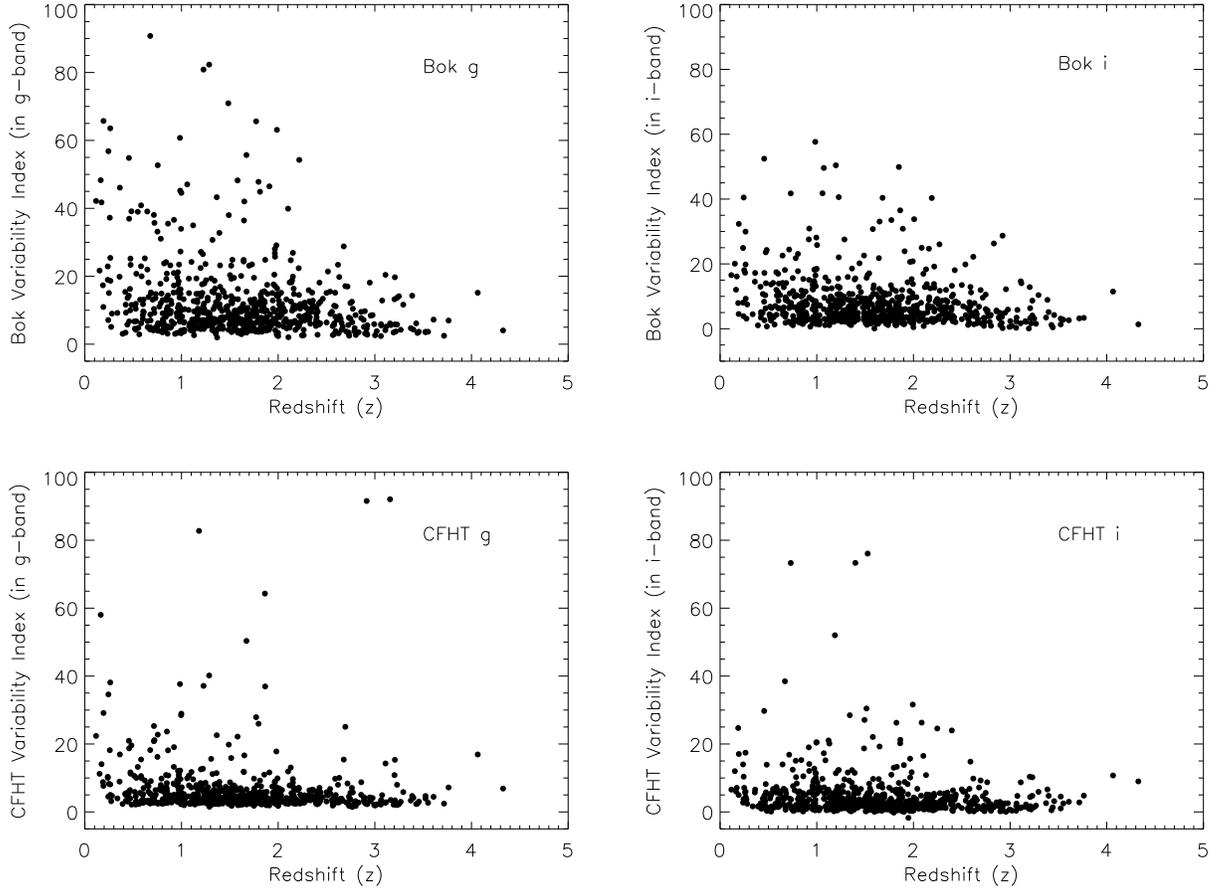}
\caption{Distribution of J-indices, constructed from $g-$ and $i-$band
  data, as a function of redshift, {\it z}.  (Upper)  J-indices were
  calculated from Bok datasets while (lower) plots have J-indices
  based from the CFHT datasets.}
\end{center}
\label{fig:jidxz}
\end{figure}

To check this conclusion, we plot the J-index against redshifts from \citet{Shen:2015}.  Figure \ref{fig:jidxz} shows
the distribution of the variability index with redshift.  There
remains a trend of larger variation in nearby AGN, but this level of
large variation is visible through the whole range of redshift.  One
of the goals of the SDSS-RM project is to study the AGN activity for a
wide range of redshifts.  The presence of different levels of
variation across all redshifts in the SDSS-RM sample confirms that reverberation mapping
should be feasible over this range with no pre-selection on quasar
variable properties.

\begin{figure}
\begin{center}
\includegraphics[scale=0.7, angle=0]{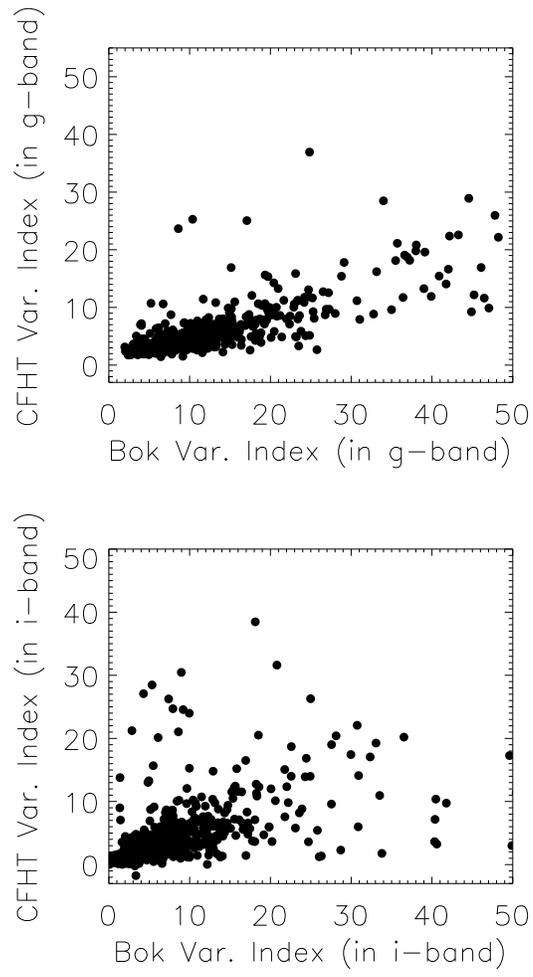}
\caption{Comparison of the Welch-Stetson J-index constructed from Bok
  and CFHT data. The upper panel is the comparison of J-index based on
the $g-$band data.  The lower panel is for the J-index based on the
$i-$band data.}
\end{center}
\label{fig:telecompare}
\end{figure}

To understand the J-index calculated from data taken at two different
telescopes, we plot the indices from the $g-$ and $i-$band data in
Figure \ref{fig:telecompare}.  There is no one-to-one correlation
between the J-indices calculated from the Bok and CFHT data.  The
reason for the typically larger J-index for Bok data could be due to
the differing  S/N of the individual images from each telescope.  The
uncertainties of the flux measurements are folded into the J-index
calculation (see Appendix for more details), and the $g-$band
uncertainties are smaller per Bok exposure.  Thus, the Bok J-index
values will appear to have a larger variation compared to the J-index
based on the CFHT datasets, for both $g-$ and $i-$ bandpasses. 
As mentioned above, the Bok dataset has more epochs and a longer time
baseline than the CFHT dataset.  Variability will be more likely to be
detected in the Bok data than with the CFHT data.
 
\begin{figure}
\begin{center}
\includegraphics[scale=0.7,angle=0]{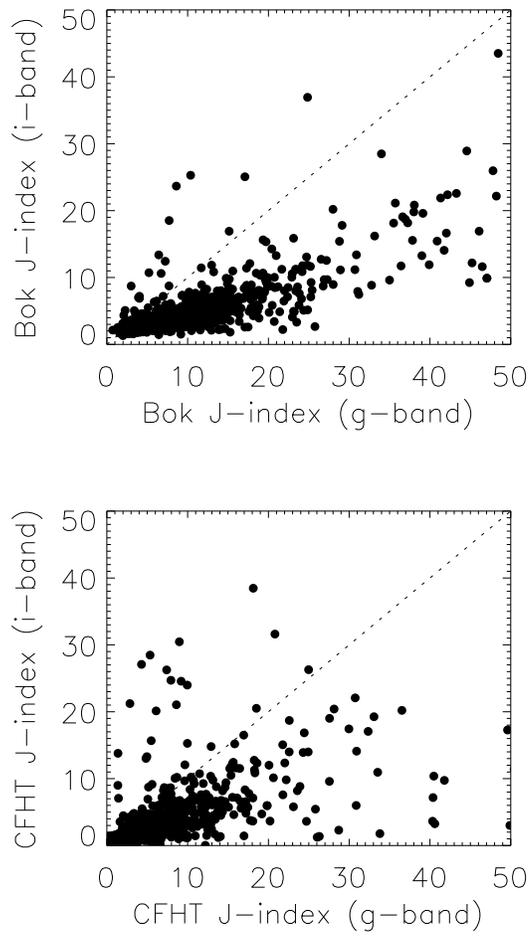}
\caption{Comparison of the J-indices constructed from $g-$ and
  $i-$band data.  A one-to-one line is provided to show that the
  quasars are more variable at shorter wavelengths.  Upper panel:
  J-index based on Bok data.  Lower panel: J-index based on CFHT data.}
\end{center}
\label{fig:idcompare}
\end{figure}

Further, we compared separately the J-index calculated based by
telescope.  In Figure \ref{fig:idcompare}, we compare the J-index
based on $g-$ and $i-$band data, and again, we do not see a clear
one-to-one correlation.  In this case, the $g-$band J-index shows a
higher variability value than the $i-$band based data.  The
variability in amplitudes are generally larger in shorter wavelengths
(e.g., \citealt{Macleod10}).  This is likely due to a combination of local accretion disk
temperature changes being more evident with shorter wavelength and
effective disk areas being smaller with shorter wavelength, the latter
leading to shorter-wavelength flux more closely tracking variations in
illuminating flux from the innermost regions of the accretion flow.
In Figure \ref{fig:idcompare}, we added the one-to-one trend line to
help guide the eye, and to emphasize how it is clearly not
one-to-one.

\section{Summary}

We present the photometric datasets for the SDSS-RM AGNs obtained at
the CFHT and Bok telescopes over a span of 4 years
(2014-2017).  We applied ISIS, a differential flux photometry
technique, to the time-series data, and we calculated a variability
measure, the Welch-Stetson J-index.  The variability index can be used to 
select interesting AGN for further study.  We find that many of the
variable AGN in this sample have long-term variability (i.e.,
the variation occurs over years rather than days).  We were able to
identify objects with low levels of variability that appear to be significant and
above the noise level.  Almost all the objects in this study appear to
have some level of variation in their continuum flux over the course
of the observations.

There are many projects currently underway using these
photometric data.  The SDSS-RM team has been merging the Bok
and CFHT datasets with the SDSS BOSS spectra to analyze the
various lags seen in the AGN (see \citealt{Grier:2017} and
\citealt{Grier:2019}).  The photometry is also being used to measure
inter-band continuum lags in quasars to learn about the accretion disk
size \citep{Homayouni:2019}.  Over 200 AGNs from the original SDSS-RM
sample have merged Bok and CFHT light curves that have been published
by \citet{Grier:2017}, and more are expected to be merged and analyzed
for future projects (e.g., \citealt{Grier:2019},
\citealt{Homayouni:2019}, and \citealt{Homayouni:2020}).

\acknowledgements

{Based in part on observations obtained with MegaPrime/MegaCam, a joint
project of CFHT and CEA/DAPNIA, at the Canada-France-Hawaii Telescope
(CFHT) which is operated by the National Research Council (NRC) of
Canada, the Institut National des Sciences de l'Univers of the Centre
National de la Recherche Scientifique of France, and the University of
Hawaii.

The authors wish to recognize and acknowledge the very significant
cultural role and reverence that the summit of Maunakea has always had
within the indigenous Hawaiian community.  We are most fortunate to
have the opportunity to conduct observations from this mountain.

We thank the CFHT staff who supported our program, including Nadine
Manset, Todd Burdullis, Simon Prunet, and Andreea Petric.

We thank the anonymous referee for the helpful comments to this paper.

PH acknowledges the support of the Natural Sciences and Engineering
Research Council of Canada (NSERC), funding reference number
2017-05983.  CJG, WNB, and DPS acknowledge support from NSF grant AST-1517113. YS
acknowledges support from an Alfred P. Sloan Research Fellowship and
NSF grant AST-1715579.  LCH acknowledges the National Key R\&D Program
of China (2016YFA0400702) and the National Science Foundation of China
(11721303, 11991052).  JVHS and KH acknowledge support from a STFC grant
ST/R000824/1. CSK is supported by NSF grants AST-1814440 and AST-1908570.
}

\appendix
\section{The Welch-Stetson J Variability Index}
The Welch-Stetson J-index (or simply the J-index) has many features that best incorporate the SDSS-RM
photometric datasets to produce a robust quantification of the AGN
variation.  The J-index takes advantage of the observing cadence,
specifically if there were paired epochs close in time.  For the Bok
data, two observations were taken of each subfield, depending on the
sky conditions.  We can also calculate the J-index for unpaired,
single epochs.  This feature allows us to use all of the
observations rather than tossing out unpaired epochs.  To define a
``pair'' of observations from a night's dataset, we selected a minimum
time interval between observations of 0.005 days (7 minutes).  Any
two observations taken within this interval were considered to be a
``pair'';  any observation taken outside of this interval was
considered to be part of another (subsequent) observation or pair.  If
an observation does not have a closely-timed observation, it was
considered to be a ``single'' observation of that subfield.  In our
datasets, we assume the second of a pair of observations was collected
within 7 minutes after the collection of the first observation.  This time
interval can be adjusted to define ``paired'' observations for a
specific dataset.  

Another strength of the J-index is that it uses information from two
different bandpasses.  For variable stars, the variation is often
correlated with color due to the nature of the stellar pulsation.
However, for AGNs, the observed variability in the light curve does
not necessarily correlate in $g$ and $i$ at the same moment in time as
time delays between the $g$ and $i-$band continuum fluxes have been observed
(e.g., \citet{Shappee:2014}, \citet{Fausnaugh:2016, Fausnaugh:2018},
 \citet{Homayouni:2019}, and \citet{Cackett:2020}).  Thus, we chose not to combine the $g$
and $i$ datasets in the construction of the J-index.

\indent
The J-index is composed of the relative error ($\delta$), normalized
residuals of a pair of observations ($P_{k}$), and
a weighting factor ($w_{k}$).   \citet{Stetson:1996} defines the relative
error as the residual from a mean magnitude, weighted by the
measurement's uncertainty and the number of observations, {\it n}, as

\begin{equation}\label{residualerror}
\delta_{i} = \frac{f_{i}-\bar{f}}{\sigma_{f,i}}\sqrt{\frac{n}{n-1}}
\end{equation}

In the case of our dataset, we do not use magnitudes, but
differential flux values from ISIS, thus in Equation \ref{residualerror}, instead of using
\citet{Stetson:1996}'s ``{\it v} magnitude'', we are using ``{\it f}'' for the
differential flux.  

As described in \citet{Stetson:1987} and \citet{Stetson:1996}, we
follow the prescription to obtain a robust weighted mean flux value
through iteration.  When the initial weighted mean flux value has been
determined from Equation \ref{residualerror}, we recalculate the
weight by its residual error.  This weight of an observation is
defined as

\begin{equation}\label{weight}
w_{i} = \Big[ 1 + \left(\frac{|\delta_{i} |}{2}\right)^2\Big]^{-1}
\end{equation}

The following equation is used for the calculation for the weighted
mean flux from \citet{Welch:1993} (their equation 1 and 2, computed in
terms of flux):

\begin{equation}\label{meanflux}
\bar{f} = \sum^{n}_{i=1} \frac{f_{i}}{\sigma^{2}_{f,i}} / \sum^{n}_{i=1} \frac{1}{\sigma^{2}_{f,i}}
\end{equation}

For our weighted mean flux, $\bar{f}$, we iterated five times to reach
a stable value and to minimize the effects of any large outliers in the
light curve data.    We found that increasing the number of
iterations did not change the value of the mean flux
significantly.  A quick visual inspection revealed that
some of the light curves did show a real, large outlier from the
photometry.  To help aid in the robustness of our variability index, we
removed any large, obvious outliers that were $3\sigma$ from our
weighted mean flux. In some cases, a single outlier did influence the
calculation of the variability index, and with $3\sigma$ clipping,
we have a more robust index value.  

For a pair of closely timed data points, the product of the residual errors
can be constructed from the {\it i}-th and {\it j}-th observations.  This product
is identified as the normalized residuals of the {\it k}-th pair

\begin{equation}\label{pair}
P_{k} = \delta_{i}\delta_{j}
\end{equation}

The normalized residual for a single observation can be
simplified to

\begin{equation}\label{singleton}
P_{k} = \delta^{2}_{i} -1 
\end{equation} 

The J-index is calculated from Equation \ref{jval}.  The \rm{sgn} function simply returns the sign of the
expression, whether it is positive or negative.  The subscript $k$
refers to the paired observation.  We use the weighting factor
determined from Equation \ref{weight} here for the J variability
index.  In this weighting factor, we use the residual error of the
pair or single observation and does not require iteration used with the
weighted mean flux.

\begin{equation}\label{jval}
J=\frac{\sum^{n}_{k=1}w_{k} sgn(P_{k}) \sqrt{|P_{k}|}}{\sum^{n}_{k=1}w_{k}}
\end{equation}

The advantage of using paired observations compared to single
observations is that the $P_{k}$ quantity will be positive in most
cases.  The individual residual errors will be positive for fluxes far
above the weighted mean or negative for fluxes far below.  A quasar
with strong variability can have $P_{k}$ values ranging from small to
large values, but the sum will produce a large J index value.  

In a few cases, Equation \ref{jval} yielded a negative value.
\citet{Stetson:1996} describes how cosmetic defects or problems with
the data processing can produce this negative value.  We identified
four AGN with a negative J-values in the Bok dataset, and upon
inspection of the reference frame, all of these objects are found near
the edge of a CCD where the PSF photometry can fail.


\clearpage

\startlongtable
\begin{deluxetable}{lclccl}

\tablewidth{0pt}

\tablecaption{Log of CFHT Observation}
\tablenum{1}
\label{cfhtlog}

\tablehead{\colhead{MJD} & \colhead{UT Date} & \colhead{Filter} &
  \colhead{Avg Seeing} & \colhead{Weather} & \colhead{Comment}  \\ 
\colhead{(days)} & \colhead{} & \colhead{} & \colhead{(arcsec)} &
\colhead{} & \colhead{} }

\startdata
\hline
2014A &            &       &       &             & \\ \hline
56712 &  20140224  &  g,i  &  0.7 & photometric & \\
56715 &  20140227  &  g,i  &  1.0 & photometric &        \\
56721 &  20140305  &  g    &  1.2 & photometric & partial coverage \\
56722 &  20140306  &  g,i  &  0.7 & photometric &  \\
56723 &  20140307  &  g,i  &  0.6 & thin cirrus & \\
56741 &  20140325  &  g,i  &  1.0 & thin cirrus &  \\
56743 &  20140327  &  g    &  0.8 & photometric & partial coverage \\
56744 &  20140328  &  g,i  &  0.7 & photometric & \\
56751 &  20140404  &  g,i  &  0.7 & photometric & \\
56772 &  20140425  &  g    &  0.9 & thin cirrus & \\
56776 &  20140429  &  g,i  &  0.6 & photometric & \\
56778 &  20140501  &  g,i  &  0.7 & photometric & \\
56780 &  20140503  &  g,i  &  0.4 & thin cirrus & \\
56781 &  20140504  &  g,i  &  0.5 & thin cirrus & partial coverage in i \\
56783 &  20140506  &  g,i  &  0.5 & partly cloudy & partial coverage in i \\
56798 &  20140521  &  g,i  &  0.6 & photometric & \\
56801 &  20140524  &  g,i  &  0.8 & photometric & \\
56803 &  20140526  &  g    &  0.9 & thin cirrus & \\
56805 &  20140528  &  g,i  &  0.7 & photometric & \\
56807 &  20140530  &  g,i  &  0.8 & photometric & \\
56809 &  20140601  &  g    &  0.8 & photometric & \\
56811 &  20140603  &  g,i  &  0.5 & photometric & \\
56829 &  20140621  &  g,i  &  0.6 & photometric & partial coverage (CFHTLS D3)\\
56831 &  20140623  &  g,i  &  0.5 & photometric & \\
56833 &  20140625  &  g,i  &  0.6 & photometric & \\
56835 &  20140627  &  g,i  &  0.6 & photometric & \\
56837 &  20140629  &  g,i  &  0.6 & photometric & \\
56839 &  20140701  &  g,i  &  0.8 & photometric & partial coverage in i \\
56841 &  20140703  &  g,i  &  0.6 & thin cirrus & \\
56865 &  20140727  &  g,i  &  0.8 & photometric & partial coverage (CFHTLS D3)\\
\hline
2015A  &            &         &       &             & \\ \hline
57078  &  20150225  &  g    &  1.0 & photometric & \\
57104  &  20150323  &  g    &  0.6 & partly cloudy & \\
57105  &  20150324  &  g    &  0.8 & photometric & short exposures only \\
57123  &  20150411  &  g    &  0.8 & thin cirrus & \\
57133  &  20150421  &  g    &  0.8 & photometric & partial coverage \\
57134  &  20150422  &  g    &  1.0 & thin cirrus & partial coverage \\
57157  &  20150515  &  g,i  &  1.0 & photometric & \\
57160  &  20150518  &  g,i  &  0.8 & thin cirrus & partial coverage in $i$ \\
57166  &  20150524  &  g    &  0.7 & photometric & \\
57185  &  20150612  &  g    &  1.0 & photometric & \\
57191  &  20150618  &  g    &  1.0 & photometric & \\
57213  &  20150710  &  g,i  &  1.0 & photometric & partial coverage in $i$ \\
57216  &  20150713  &  g,i  &  1.0 & photometric & partial coverage in $i$ \\
57217  &  20150714  &  i    &  0.9 & photometric & \\
57220  &  20150717  &  g,i  &  1.0 & thin cirrus & partial coverage in $i$ \\
57221  &  20150718  &  i    &  0.5 & photometric & \\
57223  &  20150720  &  g,i  &  0.8 & photometric & \\
\hline
2016A  &            &         &       &             & \\ \hline
57420  &  20160202  &  g,i  &  0.9 & photometric & \\
57433  &  20160215  &  g,i  &  0.6 & photometric & \\
57479  &  20160401  &  g,i  &  0.6 & photometric & \\
57492  &  20160414  &  g,i  &  1.1 & photometric & \\
57508  &  20160430  & g,i  &  0.9 & photometric & \\
57520  &  20160512  &  g,i  &  0.6 & thin cirrus & \\
57542  &  20160603  &  g    &  0.8 & photometric & \\
57543  &  20160604  &  i    &  0.6 & photometric & \\
57547  &  20160608  &  g    &  0.6 & photometric & \\
57567  &  20160628  &  g    &  0.9 & partly cloudy & \\
57568  &  20160629  &  i    &  0.9 & thin cirrus & partial coverage \\
\hline
2017A  &            &         &       &             & \\ \hline
57806  &  20170222  &  g    &  1.0 & photometric & partial coverage \\
57807  &  20170223  &  g    &  1.0 & photometric & \\
57833  &  20170321  &  g    &  0.9 & photometric & \\
57840  &  20170328  &  g    &  0.8 & photometric & \\
57843  &  20170331  &  g    &  0.7 & photometric & \\
57846  &  20170403  &  g    &  0.8 & photometric & \\
57862  &  20170419  &  g    &  0.8 & photometric & \\
57868  &  20170425  &  g    &  0.9 & partly cloudy & \\
57871  &  20170428  &  g    &  0.7 & photometric & \\
57875  &  20170502  &  g    &  1.1 & thin cirrus & \\ 
57877  &  20170504  &  g    &  0.8 & photometric & \\
57890  &  20170517  &  g    &  0.7 & photometric & \\
57894  &  20170521  &  g    &  0.6 & thin cirrus & \\
57903  &  20170530  &  g    &  0.6 & thin cirrus & \\
57924  &  20170620  &  g    &  1.0 & photometric & \\
57950  &  20170716  &  g    &  0.8 & thin cirrus & \\
57964  &  20170730  &  g    &  0.8 & photometric & \\
\enddata

\tablecomments{``Partial coverage'' indicates not all RM pointings
  were imaged that epoch. Epochs MJD=$56829$ and $56865$ were
  pointings to observe additional calibration sources in the CFHT-LS
  D3 field, which partially covers the RM field.  The Weather column
  refers to the weather status during our observations that night and
  are taken from the searchable CFHT QSO Night Reports and Program
  Status available at http://www.cfht.hawaii.edu/en/science/QSO/\#nr,
  where more than one weather condition is noted during our
  observations, we cite the worst condition reported.}

\end{deluxetable}

\clearpage

\begin{deluxetable}{lll}
\tablecaption{CFHT Pointing Centers \label{table:t_pbh1}}
\tablenum{2}
\tablewidth{0pt}
\tablehead{
\colhead{Position} &
\colhead{RA} &
\colhead{DEC}
}
\startdata
A1 & 14:14:47.17 & +53:05:46.0 \\
A2 & 14:14:50.83 & +53:03:31.0 \\
B1 & 14:14:47.19 & +52:07:00.5 \\
B2 & 14:14:50.81 & +52:04:45.5 \\
C1 & 14:08:43.77 & +52:09:50.0 \\
C2 & 14:08:47.54 & +52:07:35.0 \\
D1 & 14:08:15.56 & +53:05:46.0 \\
D2 & 14:08:19.40 & +53:03:31.0 \\
E1 & 14:08:35.77 & +54:01:42.0 \\
E2 & 14:08:39.63 & +53:59:27.0 \\
F1 & 14:14:47.15 & +54:04:31.5 \\
F2 & 14:14:50.85 & +54:02:16.5 \\
G1 & 14:20:58.53 & +54:01:42.0 \\
G2 & 14:21:02.07 & +53:59:27.0 \\
H1 & 14:21:18.77 & +53:05:46.0 \\
H2 & 14:21:22.26 & +53:03:31.0 \\
I1 & 14:20:50.61 & +52:09:50.0 \\
I2 & 14:20:54.08 & +52:07:35.0 \\
\enddata
\end{deluxetable}

\clearpage

\startlongtable

\begin{deluxetable}{ccccc}

\tablecaption{Log of Bok Observations}
\tablenum{3}
\label{bokobslog}

\tablehead{\colhead{MJD} & \colhead{UT Date} & \colhead{Filter(s)} & \colhead{Ave. Seeing} & \colhead{Comment} \\ 
\colhead{(days)} & \colhead{} & \colhead{} & \colhead{(arcsec)} & \colhead{} } 

\startdata
2014A &      &      &     &    \\ \hline
56648 & 20131222 & g & 3.0 & partial coverage \\
56671 & 20140114 & g & 1.5 &  \\
56672 & 20140115 & i & 1.7 &  \\
56673 & 20140116 & i & 1.7 &  \\
56674 & 20140117 & i & 1.4 &  \\
56675 & 20140118 & g & 1.4 &  \\
56677 & 20140120 & g & 1.5 &  partial coverage, partly cloudy\\
56678 & 20140121 & g,i & 1.2 &   \\
56680 & 20140123 & g,i & 1.7 & thin clouds \\
56683 & 20140126 & g,i & 1.4 &  \\
56685 & 20140128 & g & 1.5 &  \\
56686 & 20140129 & i & 1.3 &  cloudy\\
56701 & 20140213 & g & 1.7 &  thin clouds\\
56702 & 20140214 & i & 1.5 &  thin clouds \\
56703 & 20140215 & i & 1.4 &  \\
56705 & 20140217 & g & 1.4 & partial coverage, cloudy\\
56706 & 20140218 & g & 1.6 &  \\
56707 & 20140219 & i & 1.8 &  partly cloudy\\
56728 & 20140312 & g & 1.5 &  \\
56729 & 20140313 & i & 1.2 &  thin clouds\\
56730 & 20140314 & g & 2.0 &  cloudy\\
56731 & 20140315 & i & 1.3 &  \\
56732 & 20140316 & i & 2.5 & windy \\
56733 & 20140317 & i & 1.4 &  \\
56734 & 20140318 & g & 1.5 &  \\
56760 & 20140413 & g & 2.2 & partly cloudy with wind  \\
56761 & 20140414 & i & 1.4 &  partly cloudy with wind \\
56762 & 20140415 & g,i & 1.8 &  \\
56763 & 20140416 & i & 1.0 &  \\
56764 & 20140417 & g & 1.9 &  \\
56771 & 20140424 & g & 1.6 &  partly cloudy\\
56772 & 20140425 & i & 1.2 &  \\
56773 & 20140426 & g & 1.9 & partial coverage, cloudy \\
56774 & 20140427 & g & 1.7 &  \\
56775 & 20140428 & i & 1.5 &  \\
56789 & 20140512 & g & 2.2 &  \\
56790 & 20140513 & g & 2.1 &  \\
56791 & 20140514 & i & 2.0 &  \\
56792 & 20140515 & i & 1.3 &  \\
56793 & 20140516 & g & 1.6 &  thin clouds\\
56794 & 20140517 & g & 1.3 &  thin clouds\\
56795 & 20140518 & g,i & 1.4 &  thin clouds with wind\\
56817 & 20140609 & g & 1.4 &  \\
56818 & 20140610 & g & 1.4 &  \\
56819 & 20140611 & g,i & 1.9 &  \\
56820 & 20140612 & i & 1.1 & partly cloudy \\
56821 & 20140613 & i & 1.7 &  \\
56838 & 20140630 & g & 1.5 & \\
56839 & 20140701 & i & 1.3 & \\
56840 & 20140702 & g & 1.5 & \\
56855 & 20140717 & i & 1.3 & partly cloudy\\
56856 & 20140718 & g & 1.9 & \\
\hline
2015A  &    &   &       &     \\ \hline
57031 & 20150109 & g & 1.7 & partly cloudy \\
57032 & 20150110 & g & 2.1 & \\
57057 & 20150204 & g & 1.5 & \\
57058 & 20150205 & i & 1.5 & \\
57062 & 20150209 & g & 1.5 & \\
57063 & 20150210 & i & 1.3 & \\
57087 & 20150306 & g & 2.5 & \\
57088 & 20150307 & i & 1.7 & \\
57116 & 20150404 & g,i & 1.5 & partial coverage in $g$, cloudy \\
57117 & 20150405 & g,i & 1.8 & partly cloudy\\
57148 & 20150506 & i & 1.6 & partly cloudy\\
57149 & 20150507 & g & 2.2 & partial coverage, cloudy\\
57150 & 20150508 & i & 2.1 &  \\
57170 & 20150528 & g,i & 1.5 & \\
57171 & 20150529 & g,i & 1.9 & partial coverage in $g$ \\
57211 & 20150708 & g,i & 1.9 & partial coverage in $i$ \\
57212 & 20150709 & i & 1.4 & partial coverage \\
\hline
2016A  &    &     &    &    \\ \hline
57417 & 20160130 & g,i & 1.6 &  partly cloudy\\
57435 & 20160217 & g & 3.0 & \\
57451 & 20160304 & g & 1.1 & partial coverage, cloudy \\
57467 & 20160320 & g,i & 1.3 & cloudy\\
57480 & 20160402 & g & 1.7 & partial coverage \\
57481 & 20160403 & g & 1.7 & partial coverage, cloudy \\
57494 & 20160416 & g & 3.1 & partial coverage \\
57509 & 20160501 & g & 2.0 & partial coverage \\
57522 & 20160514 & g & 1.8 & partly cloudy \\
57523 & 20160515 & i & 2.6 & cloudy\\
57535 & 20160527 & g & 1.4 & \\
57551 & 20160612 & g & 1.4 & \\
57578 & 20160709 & g,i & 2.3 & partial coverage in $g$ \\
57579 & 20160710 & g & 2.2 & partial coverage, cloudy \\
\hline
2017A  &     &     &    &    \\ \hline
57788 & 20170204 & g & 1.5 & partial coverage, cloudy \\
57798 & 20170214 & i & 1.6 & partial coverage \\
57806 & 20170222 & g & 1.4 & partial coverage \\
57818 & 20170306 & g & 2.1 & cloudy\\
57830 & 20170318 & i & 1.2 & partial coverage \\
57831 & 20170319 & g,i & 1.2 & partial coverage in $i$, partly cloudy \\
57847 & 20170404 & g & 2.3 & \\
57848 & 20170405 & i & 1.4 & \\
57855 & 20170412 & g,i & 1.5 & \\
57895 & 20170522 & g & 1.8 & \\
57904 & 20170531 & g & 1.3 & partial coverage, cloudy \\
57905 & 20170601 & g & 2.1 & cloudy\\
57907 & 20170603 & g & 1.6 & \\
57920 & 20170616 & g & 1.7 & \\
57921 & 20170617 & i & 1.5 & \\
57960 & 20170726 & g & 1.9 & partly cloudy\\
\enddata
\end{deluxetable}

\clearpage

\startlongtable

\begin{deluxetable}{cccccccccc}

\tablecaption{SDSS-RM objects with the Welch-Stetson Variability Index in SDSS $g$ and $i$-band}

\tablenum{4}
\label{varidx}

\tablehead{\colhead{RM ID} & \colhead{$\alpha$} & \colhead{$\delta$} & \colhead{z} & \colhead{$g$} & \colhead{$i$} & \colhead{Bok $g$} & \colhead{Bok $i$} & \colhead{CFHT $g$} & \colhead{CFHT $i$} \\ 
\colhead{} & \colhead{(J2000.0)} & \colhead{(J2000.0)} &
\colhead{(kpc)} & \colhead{mag} & \colhead{mag} & \colhead{Var Idx} &
\colhead{Var Idx} & \colhead{Var Idx} & \colhead{Var Idx} } 

\startdata
0 & 213.697906 & 53.065586 & 1.4819 & 22.388 & 21.641 & 7.946 & 4.394 & 9.484 & 2.111  \\
1 & 213.654312 & 53.073048 & 1.4628 & 21.250 & 20.837 & 9.161 & 1.527 & 3.309 & 1.370   \\
2 & 213.774719 & 53.060932 & 1.2880 & 21.060 & 20.669 & 5.440 & 2.411 & 4.163 & ...   \\
3 & 213.673996 & 53.149841 & 2.2371 & 22.271 & 21.730 & 5.686 & 3.034 & 2.882 & 1.220  \\
4 & 213.756454 & 53.016392 & 1.7572 & 21.926 & 21.629 & 6.438 & 1.450 & 5.666 & 1.692   \\
5 & 213.814209 & 53.116009 & 0.7585 & 20.638 & 20.262 & ... & ... & 3.426 & 2.671   \\
6 & 213.664200 & 53.008118 & 1.2851 & 20.275 & 19.324 & 7.954 & 4.804 & 7.097 & 3.796   \\
7 & 213.785721 & 53.005531 & 2.7656 & 21.356 & 21.254 & 12.253 & 2.905 & 3.026 & 1.288   \\
8 & 213.826141 & 53.147217 & 1.8474 & 22.581 & 21.796 & 3.912 & 7.003 & 2.517 & 4.005   \\
9 & 213.723190 & 53.180897 & 1.3278 & 22.179 & 21.662 & 6.740 & 2.799 & 7.635 & 2.777  \\
\enddata
\tablecomments{This table is published in its entirety in the
  electronic version of this paper and is available in a machine
  readable form.  A portion is provided here to guide in its form and content.}
\end{deluxetable}

\end{document}